\documentclass{article}
\usepackage{spconf, graphicx, amsfonts, mathtools}
\usepackage{amssymb,bm}
\usepackage{amsmath,bbm}
\usepackage{epstopdf,multirow,tabularx}

\title{Reinforcement Learning Based Speech Enhancement for Robust Speech Recognition}
\name{Yih-Liang Shen\textsuperscript{1}, Chao-Yuan Huang\textsuperscript{1}, Syu-Siang Wang\textsuperscript{2}, Yu Tsao\textsuperscript{3} , Hsin-Min Wang\textsuperscript{4} and Tai-Shih Chi\textsuperscript{1}}
\address{\textsuperscript{1}Department of Electrical and Computer Engineering, National Chiao Tung University, Hsinchu, R.O.C\\
	\textsuperscript{2} Joint Research Center for AI Technology and All Vista Healthcare, MOST, Taipei, R.O.C\\
	\textsuperscript{3} Research Center for Information Technology Innovation, Academia Sinica, Taipei, R.O.C\\
	\textsuperscript{4} Institute of Information Science, Academia Sinica, Taipei, R.O.C}

\begin{document}
	\ninept
	\maketitle
		\begin{abstract}	
		Conventional deep neural network (DNN)-based speech enhancement (SE) approaches aim to minimize the mean square error (MSE) between enhanced speech and clean reference. The MSE-optimized model may not directly improve the performance of an automatic speech recognition (ASR) system. If the target is to minimize the recognition error, the recognition results should be used to design the objective function for optimizing the SE model. However, the structure of an ASR system, which consists of multiple units, such as acoustic and language models, is usually complex and not differentiable. In this study, we proposed to adopt the reinforcement learning algorithm to optimize the SE model based on the recognition results. We evaluated the propsoed SE system on the Mandarin Chinese broadcast news corpus (MATBN). Experimental results demonstrate that the proposed method can effectively improve the ASR results with a notable $12.40\%$ and $19.23\%$ error rate reductions for signal to noise ratio at 0 dB and 5 dB conditions, respectively.
	\end{abstract}
	\vspace{-0.1cm}
	\begin{keywords}
		reinforcement learning, automatic speech recognition, speech enhancement, deep neural network, character error rate
	\end{keywords}
	\vspace{-0.4cm}
	\section{Introduction}
	The performance of automatic speech recognition (ASR) has significantly improved in recent years. However, a long-existing issue still remains: ASR suffers severe performance degradation in noise environments \cite{li2015robust}. Many approaches have been proposed to address the noise issue. One category of these approaches is speech enhancement (SE) \cite{li2013investigation,wang2016joint}. The goal of SE is to generate enhanced speech signals that closly match clean and undistorted speech signals, by removing the noise components from the noisy speech \cite{acero1990environmental,hsieh2016employing,zhang2018training}. Traditional SE approaches are designed based on some assumptions of speech and noise characteristics \cite{cohen2002noise,mcaulay1980speech}. Generally, these approaches can yield a satisfactory performance in terms of speech quality but may not be directly beneficial in the improvement of the ASR performance \cite{du2014robust,hsieh2016employing}. 
	
	Recently, deep-learning-based SE approaches have received increased attention and it has been confirmed that they yield better performances than traditional methods in many tasks \cite{baby2015exemplar,DBLP:journals/corr/Simpson15e,wang2018supervised}. Because of the deep structure, the deep-learning-based models can effectively characterize the complex transformation of noisy speech to clean speech, or they can precisely estimate a mask to filter out noise components from the noisy speech. To train the deep-learning-based models, the mean square error (MSE)-criterion is usually used as the objective function. Specifically, the model is trained to minimize the MSE of the enhanced speech and clean references. Although it has been proven that the MSE-based objective function is effective for noise reduction, it is not optimal for improving speech quality and intelligibility, or the ASR performance \cite{xu2015regression,lu2013speech,meng2018adversarial,meng2018cycle}.  
	
	Clearly, the ASR results should be the optimal objective function for SE. However, most of the commonly used ASR systems consist of multiple modules, such as the acoustic models and language models. Correspondingly, the input--output correlation is extremely complicated and may not be differentiable. Thus, it is difficult to directly use the recognition results to directly optimize the SE models. Moreover, it takes a considerable amount of resources to build an ASR system, and thus the use of a well-established ASR system from a third party is thus favorable. In this study, we propose to adopt the reinforcement learning (RL) algorithm to train an SE model to minimize the recognition errors. 
	
	The main concept of the RL algorithm is to take an action in an environment in order to maximize some notion of a cumulative reward \cite{sutton1992reinforcement}. Different from supervised and unsupervised learning algorithms, the RL algorithm learns how to attain a (complex) goal in an iterative manner. 
	To-this-date, the RL algorithms have been successfully applied to various tasks, such as robot control \cite{kohl2004policy}, dialogue management \cite{singh2002optimizing}, and computer game playing \cite{mnih2015human}. 
	
	The RL algorithm has also been adopted into the speech signal processing filed. In \cite{kala2018reinforcement}, the RL has been used to improve the ASR performance. Based on hypothesis selection by the users, the system can improve the recognition accuracy as compared to unsupervised adaptation. Meanwhile, the RL has been used for DNN-based source enhancement by optimizing objective sound quality assessment score \cite{koizumi2017dnn}. The results show that by using the RL algorithm, both perceptual evaluation of the speech quality (PESQ) \cite{rix2001perceptual} and the short-time intelligibility measure (STOI) \cite{taal2011algorithm}  scores can be improved as compared to the MSE-based training criterion \cite{koizumi2018dnn}. 
	
	In this study, we adopt the same idea presented in \cite{koizumi2017dnn} to establish an RL-based SE system to optimize the ASR performance. Instead of estimating the ratio masking as used in \cite{koizumi2017dnn}, the proposed SE system determines the optimal binary mask to minimize the recognition errors. Notably, the ASR system is fixed in the proposed method. This is to simulate most realistic scenarios that a well-trained ASR system is provided by a third party, and an SE is built to generate suitable inputs to the ASR system. We evaluated the proposed RL-based SE system on a Mandarin Chinese broadcast news corpus (MATBN) \cite{wang2005matbn}. According to our experimental results, the proposed RL-based SE system effectively decreases the character error rate (CER) during the testing of the recognition in the presence of noise. The remainder of this paper is organized as follows. Section 2 review relative techniques. Section 3 introduces the proposed system. Section 4 presents the experimental setup and results. Finally, section 5 provides conclusion remarks.
	\vspace{-0.3cm}
	\section{Related Works}
	In the time domain, a noisy speech signal $\mathbf{y}$ is formulated by a combination of a clean speech signal $\mathbf{s}$ and an additive noise signal $\mathbf{n}$. By performing short-time Fourier transform (STFT), log--power operation, and mel--frequency-based filtering, the mel--frequency power spectrogram (MPS) of $\mathbf{y}$ can be expressed as:
	\begin{equation}
	\mathbf{Y}=\mathbf{S}+\mathbf{N}.
	\end{equation}
	In this study, $p$ frames of the STFT MPS feature vectors are concatenated to form one chunk vector for  $\mathbf{Y}$, $\mathbf{S}$ and $\mathbf{N}$. Accordingly, we thus have:
	\begin{equation}
	\hat{\bm{\mathcal{X}}}_c=[\bm{\mathcal{X}}_{cp}^\top,\bm{\mathcal{X}}_{cp+1}^\top,\cdots,\bm{\mathcal{X}}_{(c+1)p-1}^\top]^\top,\:\:\bm{\mathcal{X}}\in\{\mathbf{Y},\mathbf{S},\mathbf{N}\},
	\end{equation}
	where $c=\{0,1,\cdots,C\}$ is a chunk index, and $C$ is the total number of chunks vectors within $\bm{\mathcal{X}}$. Note that when $p=1$, the chunk vector is the STFT MPS feature vector.

	\subsection{Ideal Binary Mask-based SE System}
	It has been reported that when the goal is to improve the ASR performance, ideal binary mask (IBM) is more suitable than ideal ratio mask (IRM) or directly mapping \cite{moore2017speech} to be used to design the SE system. Therefore, we implement an IBM-based SE system in this study. 
	For the IBM-based SE system, the input $\hat{\mathbf{Y}}$ was filtered by IBM to obtain the enhanced output $\hat{\mathbf{S}}'$:
	\begin{equation}
	\hat{\mathbf{S}}'=\hat{\mathbf{Y}}.\times\hat{\mathbf{B}},
	\end{equation}
	where ``$.\times$'' represents an element-wise multiplier, and $\hat{\mathbf{B}}$ is the IBM matrix, which is defined as:
	\begin{equation}\label{eq:ibm}
	\hat{\mathbf{B}} = \mathbbm{1}\{log(\hat{\mathbf{S}})-log(\hat{\mathbf{N}})\},
	\end{equation}
	where $\mathbbm{1}\{\cdot\}$ is the unit step function applied to each element of $\hat{\mathbf{B}}$.
	
	\subsection{DNN-based SE Model with the MSE Criterion}\label{sec:convdnn}
	For the DNN-based SE, a set of noisy-clean  training pairs are prepared as the input and reference of a DNN model. For the noisy $\hat{\mathbf{Y}}$, $F$ chunk vectors are then cascaded to include more context information: $\tilde{\mathbf{Y}}_c=[\hat{\mathbf{Y}}_{c-F+1}^\top,\hat{\mathbf{Y}}_{c-F+2}^\top,\cdots,\hat{\mathbf{Y}}_{c}^\top]^\top$ . The mapping process of a feedforward DNN with $L$ hidden layers is the formulated as,
	\begin{equation}\label{eq:forward}
	\begin{array}{c}
	h_1(\tilde{\mathbf{Y}}_c)=\sigma\{\mathbf{W}_1log(\tilde{\mathbf{Y}}_c)+\mathbf{b}_1\},\\
	\vdots\\[0.1cm]
	h_{\ell}(\tilde{\mathbf{Y}}_c)=\sigma\{\mathbf{W}_{\ell}h_{\ell-1}(\tilde{\mathbf{Y}}_c)+\mathbf{b}_{\ell}\},\\
	\vdots\\[0.1cm]
	h_{L}(\tilde{\mathbf{Y}}_c)=\sigma\{\mathbf{W}_{L}h_{L-1}(\tilde{\mathbf{Y}}_c)+\mathbf{b}_{L}\},\\
	\hat{\mathbf{S}}''_c=\sigma_1\{\mathbf{W}_{L+1}h_{L}(\tilde{\mathbf{Y}}_c)+\mathbf{b}_{L+1}\},
	\end{array}
	\end{equation}
	where $\mathbf{W}_{\ell}$ and $\mathbf{b}_{\ell}$ are the weight matrices and bias vectors,respectively. Both $\sigma\{\cdot\}$ and $\sigma\{\cdot\}$ are activation functions, in which $\sigma\{\cdot\}$ is the sigmoid function while $\sigma_1\{\cdot\}$ represents a linear transformation. When the MSE is used as the cost function, the parameter set $\Theta$ that consists of all of $\mathbf{W}_{\ell}$ and $\mathbf{b}_{\ell}$ in Eq. \eqref{eq:forward} is estimated by,
	\begin{equation}
	\Theta^*=\mbox{{\small$\mathop{\arg\min}_{\Theta}(\frac{1}{C}\sum_{c=1}^C\parallel log(\hat{\mathbf{S}}_c)-log(\hat{\mathbf{S}}''_c)\parallel_2^2)$}}\\
	\end{equation}
	\vspace{-0.3cm}
	\section{proposed method}
\begin{figure}[t]
		\centering
		\centerline{\includegraphics[width=0.9\columnwidth]{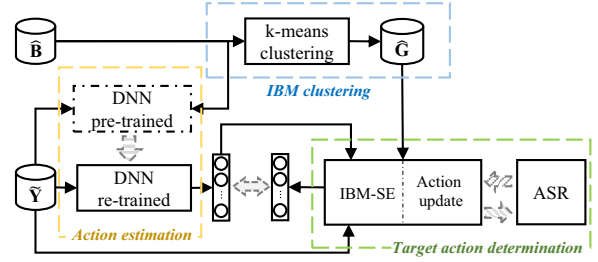}}
		\vspace{-0.3cm}\caption{The block diagram of the proposed SE system, which includes  ``\textit{IBM clustering}'', ``\textit{Action estimation}'', and ``\textit{Target action determination}''.}\label{fig:blockdiagram}	
	\end{figure}

	Figure \ref{fig:blockdiagram} illustrates the proposed system, which consists of three modules:  ``\textit{IBM clustering}'', ``\textit{Action estimation}'', and ``\textit{Target action determination}''.
	
	\subsection{IBM clustering module}
	In the IBM-based SE system, an IBM filter is computed for each feature vector. The IBM clustering module groups the entire set of IBM vectors $\hat{\mathbf{B}}$ collected from the training data to $A$ clusters based on the K-means algorithm. Each cluster is represented as $\hat{\mathbf{g}}_a$ with respect to the cluster index $a$. The ensemble of these clusters is denoted as $\hat{\mathbf{G}}$. Thus, we have,
	\begin{equation}\label{eq:kmean}
	\hat{\mathbf{G}}=[\hat{\mathbf{g}}_1,\cdots,\hat{\mathbf{g}}_a,\cdots,\hat{\mathbf{g}}_A].
	\end{equation}
	Since the elements in each IBM vector acquire binary values, the Hamming distance \cite{norouzi2012hamming} is used to compute the distance between the two vectors in this study. Meanwhile, we used 32 clusters $(A=32)$ to group $\hat{\mathbf{B}}$ based on the k-means algorithm.
	
	\subsection{Action estimation module}
	To effectively use the training data, we first pre-train the DNN model by placing $\tilde{\mathbf{Y}}_c$ at the input and  $\hat{\mathbf{B}}_c$ at the output. This pre-trained model was then re-trained with additional hidden layers to compute the $A$-dimensional action vector $\mathbf{a}''_c$ at $c$th chunk. Among the $A$ elements in $\mathbf{a}''_c$, the index with the maximum value was determined,
	\begin{equation}\label{eq:invq}
	a_c=\arg\max_{a\in\mathbb{A}}[\mathbf{a}''_c]_a,
	\end{equation}
	where $[\cdot]_{a}$ represents the $a$th element of the vector, and $\mathbb{A}=\{1,2,\cdots,A\}$. In addition, different from the spectral mapping in Eq. \eqref{eq:forward}, the softmax operation is used in the final layer in the re-trained DNN. The cost function for the re-training process is expressed as,
	\begin{equation}\label{eq:dnnsecost}
	\Theta^*=\mbox{{\small$\mathop{\arg\min}_{\Theta}(\frac{1}{C}\sum_{c=1}^C\parallel \mathbf{a}_c-\mathbf{a}''_c)\parallel_2^2)$}},
	\end{equation}
	where $\mathbf{a}_c$ is the reference target, which is derived from \textit{Target action determination} module and is described in the next section.
	
	\subsection{Target action determination module}
\begin{figure}[t]
		\centering
		\centerline{\includegraphics[width=0.8\columnwidth]{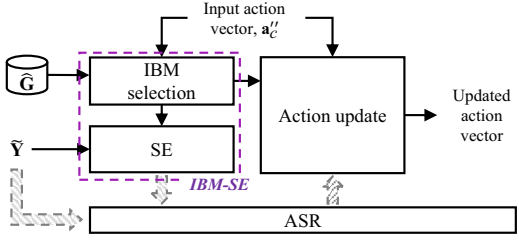}}
		\vspace{-0.3cm}\caption{The flowchart of  \textit{Target action determination} module, which is used to update the input action vector.}\label{fig:rlasr}	\vspace{-0.3cm}
	\end{figure}

	Figure \ref{fig:rlasr} shows the flowchart of the \textit{Target action determination} module. First, $\mathbf{a}_c''$, which is estimated from the action estimation module is used to determine the cluster index $a_c$ in Eq. \eqref{eq:invq}. Then, the IBM selection function selects $\mathbf{g}_{a}$ from $\hat{\mathbf{G}}$ with respect to index $a=a_c$. Next the SE function uses the selected $\mathbf{g}_{a}$ to enhance the input $\tilde{\mathbf{Y}}_c$. After enhancing all $C$ chunk vectors, both the input noisy and the IBM-enhanced STFT--MPS features are reconstructed back to the time domain signals, and then provide the ASR to calculate the utterance-based error rates (ERs), $z_{\mathbf{y}}$ and $z_{\mathbf{s}'}$, respectively. Both $z_{\mathbf{y}}$ and $z_{\mathbf{s}'}$ are used in the \textit{Target action determination} function, which is a two-stage operations, namely, the reward calculation and action update.
	
	\subsubsection{Reward calculation}
	Rather than directly use $z_{\mathbf{s}'}$ as the reward, we applied the relative value between $z_{\mathbf{y}}$ and $z_{\mathbf{s}'}$ in Eq. \eqref{eq:hptan} to avoid external factors, such as the variation of an ASR system and environmental noises. 
	\begin{equation}\label{eq:hptan}
	R = \tanh\{\alpha(z_{\mathbf{y}}-z_{\mathbf{s}'})\},
	\end{equation}
	where $\alpha > 0$ is a scalar factor, which is set to 10 in this study. For this equation, the positive $R$ denotes a larger ER of $z_{\mathbf{y}}$ than that of $z_{\mathbf{s}'}$, thus suggesting that the enhanced speech can provide better recognition results. On the other hand, a negative $R$ denotes a smaller $z_{\mathbf{y}}$ than $z_{\mathbf{s}'}$, suggesting that the enhanced speech gives worse recognition performance.
	
	In addition to the utterance-based rewards $R$, we also consider a chunk-based reward because the action for each chunk vector may act and contribute differently to $z_{\mathbf{s}'}$. That is, an effective enhancement can cause positive contribution on the ASR performance. Therefore, we defined a time-varied reward $r_c$ as:
	\begin{equation}\label{eq:weiraw}
	\hat{E}_c =(log(\hat{\mathbf{S}}_c)-log(\hat{\mathbf{S}}_c'))^\top(log(\hat{\mathbf{S}}_c)-log(\hat{\mathbf{S}}_c')),
	\end{equation}
	\begin{equation}\label{eq:wei}
	\tilde{E}_c = \frac{\hat{E}_c}{\max_{0\leq c\leq C-1}(\hat{E}_c)},
	\end{equation}
	\begin{equation}\label{eq:raw}
	r_c = \left\{ \begin{array}{lr}
	(1 - \tilde{E}_c)R,&R>0,\\[0.2cm]
	\tilde{E}_cR,&R\leq 0.
	\end{array} \right.
	\end{equation}
	From Eqs. \eqref{eq:weiraw}-- \eqref{eq:raw}, the weighting factor $\tilde{E}_c,\:\:0\leq\tilde{E}_c\leq 1$,  at the $c$th chunk is the normalized square error. When selecting a erroneous IBM vector, the normalized error $\tilde{E}_c$ in\eqref{eq:wei} is large, and accordingly $r_c$ is small, which penalizes this wrong action, as to be introduced in the next sub-section. 
	
	\subsubsection{Action update}
	To update the action vector, $\mathbf{a}_c''$, we first determine two different action indices, $a_{\hat{\mathbf{B}}_c}$ and $a_c$. To obtain $a_{\hat{\mathbf{B}}_c}$, we first follow Eq. \eqref{eq:ibm} to determine an IBM vector, which is then used to locate the closest cluster in $\hat{\mathbf{G}}$; the located cluster index is $a_{\hat{\mathbf{B}}_c}$. On the other hand, the cluster index $a_c$ is determined by Eq. \eqref{eq:invq}, as presented in the \textit{Action estimation} module.  
	
	With the determined action indices $a_{\hat{\mathbf{B}}_c}$ and $a_c$, the input action vector $\mathbf{a}_c''$  is updated for the output $\mathbf{a}_c$ based on the following equations:
	\begin{equation}
	[\mathbf{a}_c]_{a_c}= \left\{ 
	\begin{array}{*{20}{l}r}
	{r_c + max_{a_c\in\mathbb{A}}[\mathbf{a}_c'']_{a_c} },&R>0,\\
	{[\mathbf{a}_c'']_{a_c}},&R=0,
	\end{array} \right.
	\end{equation}
	and
	\begin{equation}
	[\mathbf{a}_c]_{a_{\hat{\mathbf{B}}_c}}=
	\begin{array}{*{20}{l}r}
	{[\mathbf{a}_c'']_{a_{\hat{\mathbf{B}}_c}}}-r_c,&R<0.
	\end{array}
	\end{equation}

	\subsection{Testing procedure}
	\begin{figure}[t]
		\centering
		\centerline{\includegraphics[width=0.9\columnwidth]{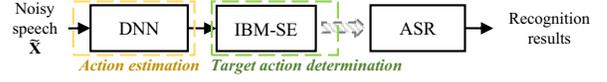}}
		\vspace{-0.3cm}\caption{The block diagram of testing part for the proposed algorithm.}\label{fig:test}	\vspace{-0.3cm}
	\end{figure}
	After performing the training on DNN with the associated objective function in Eq. \eqref{eq:dnnsecost}, Fig. \ref{fig:test} illustrates the block diagram of the testing process. From the figure, the well-trained DNN model is applied on a noisy STFT--MPS $\hat{\mathbf{X}}$, which is first extracted from the time-domain signal $\mathbf{x}$. The estimated IBM indices are then used in combination with Eq. \eqref{eq:invq} for each chunk to further enhance the input noisy and provide $\hat{\mathbf{S}}_\mathbf{x}$ in the output of the IBM--SE function. The waveform $\mathbf{s}_\mathbf{x}'$ is reconstructed from $\hat{\mathbf{S}}_\mathbf{x}$, and is then applied to ASR to conduct the recognized process.
	\vspace{-0.4cm}
	\section{EXPERIMENTS}
	\subsection{Experimental setup}
	We conducted our experiments on the MATBN task, which was an 198-hour Mandarin Chinese broadcast news corpus \cite{wang2005matbn}. The utterances in MATBN were originally recorded at a 44.1 kHz sampling rate and were further down-sampled to 16 kHz. A 25-hour gender-balanced subset of the speech utterances was used to train aset of CD-DNN-HMM acoustic models. A set of trigram language models was trained on a collection of text news documents published by the Central News Agency (CNA) between 2000 and 2001 (the Chinese Gigaword Corpus released by LDC) with the SRI Language Modeling Toolkit \cite{katz1987estimation}. The overall ASR system was implemented on the Kaldi \cite{povey2011kaldi} toolbox. Each speech waveform  was parameterized into a sequence of 40-dimensional filter-bank features. The DNN structure for the acoustic models was consisted of six hidden layers, and each layer had 2048 nodes. The dimensions for the input and output layers were 440 ($40\times(2\times5+1)$) and 2596, respectively \cite{wang2018suppression}. The evaluated results are reported as the average CER. To train the RL--SE system, another 460 utterances were selected from the MATBN corpus. The overall RL--SE and ASR systems were evaluated using another 30 utterances from the MATBN testing set. In this study, we used the baby-cry noise as the background noise.	The baby-cry noise waveform was divided into two parts, the first part was artificially added to the 460 training utterances with signal-to-noise ratio (SNR) level at 5 dB; the second part was artificially added to the 30 testing utterances at 0 and 5 dB SNR levels. Notably, the training and testing utterances were simulated using different segments of the noise source waveform, and thus the properties were slightly different. Finally, we have prepared 460 noisy--clean pairs to train the RL-based SE system. For all of the training and testing data, the applied  frame size and the shift for STFT were 32 and 16 ms in length, respectively. The 64-dimensional MPS features were then extracted from all noisy and clean utterances. Next, we established two RL-based SE models, with two different parameters $p$ for the chunk vectors: the systems with $p=1$ and $p=2$ are termed $RLSE_1$ and $RLSE_2$, respectively. Both $RLSE_1$ and $RLSE_2$ were composed of one hidden layer with 64 nodes, and 32 for the output nodes. The input dimensions of $RLSE_1$ was 704 ($64\times 1\times 11$), and that of $RLSE_2$ was 640 ($64\times 2\times 5$), in which the $11$ and $5$ are values of the parameter $F$, and is used for providing the context information (as mentioned in Section \ref{sec:convdnn}).
	
	\subsection{Experimental results}
	\begin{figure}[t]
		\centering
		\centerline{\includegraphics[width=0.9\columnwidth]{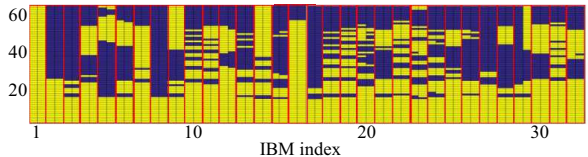}}
		\vspace{-0.3cm}\caption{Clustered IBMs were derived by the k-means algorithm.}\label{fig:ibm}	\vspace{-0.3cm}
	\end{figure}
	
	\begin{table}[b]
		\begin{center}
			\vspace{-0.3cm}\caption{The average CERs of Noisy (the baseline), $1nnSE$,  $RLSE_1$, and $RLSE_2$ at 0 and 5 dB SNR conditions.}
			\label{tab:recog}
			\begin{tabular}{c|cccc}
				\hline\hline		
				\textit{SNR}&\textit{Noisy}&\textit{$1nnSE$}&$RLSE_1$&$RLSE_2$\\
				\hline
				\hline
				\textit{5 dB }&56.14&73.09&55.60&\textbf{49.18}\\ 
				\hline
				\textit{0 dB}&81.40&85.79&77.20&\textbf{65.75}\\ 
				\hline\hline
			\end{tabular}
		\end{center}
	\end{table}
	
	Figure. \ref{fig:ibm} shows all the 32 IBM vectors, each with 64-dimensions. The IBM in Eq. \eqref{eq:kmean} used in the $RLSE_2$ system. Bright yellow elements in the figure denote ones (in terms of their binary values) and the blue elements denote zeros. From the figure, we observe that low-dimensional MPS features are dominated by speech components. One possible explanation is that the noise signals did not mask the human speech in the low-frequency regions. In addition, the entire first column consisted of ones, thus suggesting that the silence frames were also contained in the baby-cry noise.
	
	We then compared the averaged CER results of the $RLSE_1$ and $RLSE_2$ systems, and the corresponding results are listed in Table \ref{tab:recog}. The unprocessed noisy speech was also recognized by an ASR system, and the corresponding results are denoted as ``Noisy''. 	To test the effectiveness of RL learning, we designed another set of experiments: the same 32 IBM vectors were used, while the one-nearest-neighbor ($1nnSE$) method was used to determine the IBM vector for enhancement. The enhanced speech was then recognized by the same ASR system; the corresponding results were denoted as $1nnSE$ in Table \ref{tab:recog}.  
	
	When the recognition was tested using the original clean testing utterances, the CER was $11.50\%$. However, as shown in Table \ref{tab:recog}, when there was noise involved in the background, the CER was dropped considerably to $56.14\%$ and $81.40\%$, respectively, for 5 dB and 0 dB SNR levels. We then noted that $1nnSE$ could not provide any improvements over Noisy, thus showing that the one-nearest-neighbor method could not select the optimal IBM vectors for SE to improve the ASR performance. Furthermore, both $RLSE_1$ and $RLSE_2$ provided better recognition results than those of Noisy and $1nnSE$, and $RLSE_2$ outperformed $RLSE_1$. The relative CER reductions of $RLSE_2$ over Noisy are $12.40\%$ (from $56.14\%$ to $49.18\%$) at the 5 dB SNR level, and $19.23\%$ (from $81.40\%$ to $65.75\%$) for the 0 dB SNR level. The results in Table \ref{tab:recog} clearly demonstrate the effectiveness of RL-based SE for improving ASR performance in the presence of noise.	
	
	\begin{figure}[t]
		\centering
		\centerline{\includegraphics[width=0.77\columnwidth]{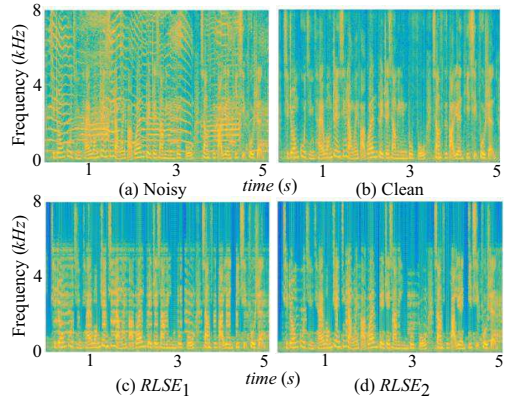}}
		\vspace{-0.3cm}\caption{The spectrograms of (a) Noisy speech, (b) clean speech, (c) enhanced speech by $RLSE_1$, and (d) enhanced speech by $RLSE_2$. }\label{fig:spec}	\vspace{-0.3cm}
		
	\end{figure}

	\begin{table}[b]
		\begin{center}
			\vspace{-0.3cm}\caption{The STOI and PESQ scores of $RLSE_1$, $RLSE_2$, and Noisy at 0 and 5 dB SNR conditions. }	\label{tab:stoi}
			\begin{tabularx}{\columnwidth}{c|>{\centering}m{0.63cm}>{\centering}m{0.9cm}>{\centering}m{0.9cm}|>{\centering}m{0.63cm}>{\centering}m{0.9cm}c}
				\hline\hline
				\multirow{2}{*}{\textit{SNR}}&\multicolumn{3}{c|}{STOI}&\multicolumn{3}{c}{PESQ}\\
				\cline{2-7}	
				&\textit{Noisy}&$RLSE_1$&$RLSE_2$&\textit{Noisy}&$RLSE_1$&$RLSE_2$\\
				\hline
				\textit{5 dB }&0.82&0.82&\textbf{0.86}&1.85&1.67&\textbf{1.96}\\ 
				\hline
				\textit{0 dB }&0.74&0.77&\textbf{0.81}&1.45&1.42&\textbf{1.59}\\ 
				\hline\hline
			\end{tabularx}
		\end{center}
	\end{table}
	
	To visually analyze the effect of the derived RL-based SE system, we presented the spectrograms of one noisy utterance at the 5 dB SNR level (as shown in Fig. \ref{fig:spec} (a)), as well as its clean and enhanced versions by $RLSE_1$ and $RLSE_2$ (as shown in Fig. \ref{fig:spec} (b), (c), and (d), respectively). From the figure, noise components of noisy datasets were effectively removed  by $RLSE_1$ and $RLSE_2$, thus showing that despite the fact that the goal was to improve the ASR performance, the RL-based SE also performed denoising on the input speech.
	
	Recent studies have reported a positive correlation between objective intelligibility scores and ASR performance \cite{moore2017speech,Xia2017UsingOR}.
	In Table \ref{tab:stoi}, we show the STOI and PESQ scores of enhanced speech processed by $RLSE_1$ and $RLSE_2$ at SNR levels of0 and 5 dB. The results of the unprocessed noisy speech, shown as Noisy, are also listed for comparison. From this table, we show that both $RLSE_1$ and $RLSE_2$ elicit higher STOI scores than Noisy and $RLSE_2$ provides again clear improvements over $RLSE_1$. From Tables \ref{tab:recog} and \ref{tab:stoi}, we can clearly note positive correlations between the STOI scores and ASR performances.
	As for the PESQ scores, $RLSE_2$ outperformed Noisy but $RLSE_1$ slightly underperformed Noisy. It can be noted that the correlation of the PESQ scores with ASR results is not as strong as that of the STOI scores and the ASR results.  
	
	\vspace{-0.3cm}
	\section{Conclusion}
	In this study, we present an RL-based SE for robust speech recognition without retraining the ASR system. By using the  recognition errors as the objective function, the RL-based SE can effectively reduce CERs by $12.40\%$ and $19.23\%$ at 5 and 0 dB SNR conditions, respectively. We also noted that although the objective is to improve ASR performance, the enhanced speech presented denoised properties and was with improved STOI scores. This study serves as a pioneering work for building an SE system with the aim to directly improve ASR performance. The designed scenario is practical in many real-world applications where an ASR engine is supplied by a third-party.
	In the future work, more noise types and SNR levels will be considered to build the RL-based SE system.

	
	\bibliographystyle{ieeetr}
	\bibliography{ref}

\end{document}